\documentclass[preprint2]{aastex}
\usepackage{amssymb}
\usepackage{amsmath}
\usepackage{graphicx}
\usepackage{lscape}
\usepackage{rotating,color}
\usepackage{color}

\shorttitle{infall model vs observations}
\shortauthors{Chang et al.}

\begin{document}


\title { The mass-dependant star formation histories of disk galaxies: infall model versus observations }

\author{R. X. Chang \altaffilmark{1}; J. L. Hou \altaffilmark{1}; S. Y. Shen \altaffilmark{1};
C. G. Shu \altaffilmark{1,2}}

\altaffiltext{1} {Key Laboratory for Research in Galaxies and Cosmology, 
Shanghai Astronomical Observatory, Chinese Academy of Sciences, 80 Nandan Road, Shanghai, China, 200030 }
 \email{crx@shao.ac.cn}
\altaffiltext{2} {Shanghai Normal University, 100 Guilin Road, Shanghai, China, 200234}

\begin{abstract}

We introduce a simple model to explore the star formation histories of
disk galaxies. We assume that the disk origins and grows by continuous 
gas infall. The gas infall rate is parametrized by the Gaussian formula 
with one free parameter: infall-peak time $t_p$. The Kennicutt 
star formation law is adopted to describe how much cold gas turns into 
stars. The gas outflow process is also considered in our model. We find that, at 
given galactic stellar mass $M_*$, model adopting late infall-peak 
time $t_p$ results in blue colors, low metallicity, high specific star 
formation rate and high gas fraction, while gas outflow rate mainly
influences the gas-phase metallicity and star formation efficiency mainly 
influences the gas fraction. Motivated by the local observed
scaling relations, we construct a mass-dependent model by
assuming low mass galaxy has later infall-peak time $t_p$ and larger 
gas outflow rate than massive systems. It is shown that this model 
can be in agreement with not only the local observations, but also 
the observed correlations between specific star formation rate and 
galactic stellar mass $SFR/M_* \sim M_*$ at intermediate redshift $z<1$. 
Comparison between the Gaussian-infall model and exponential-infall 
model is also presented. It shows that the exponential-infall 
model predicts higher star formation rate at early stage 
and lower star formation rate later than that of Gaussian-infall. 
Our results suggest that the Gaussian infall rate may be more 
reasonable to describe the gas cooling process than the 
exponential infall rate, especially for low-mass systems. 
\end{abstract}

\keywords{Galaxies:evolution -- Galaxies:photometry -- Galaxies:
stellar content}

\section{Introduction}

More than thirty years ago, Visvanathan \& Griersmith (1977) found
that early-type spiral galaxies (S0/a to Sab) exhibit optical
color-magnitude relation (CMR), that is, luminous spirals show
redder colors than less-luminous systems.

Later, many studies confirmed this phenomena and extended the CMRs of
spirals from optical to infrared bands (Tully et al. 1982; de Jong
1996; Mobasher et al. 1986; Tully et al. 1998; Peletier \& de Grijs
1998; Bell \& de Jong 2000; Blanton et al. 2003; Baldry et al. 2004;
Macarthur et al. 2004). Chang et al. (2006a) collected a sample of
late-type galaxies from Sloan Digital Sky Survey (SDSS) and
Two-Micron All-Sky Survey (2MASS) and found that, after correcting
the effect of dust attenuation, the optical-infrared (optical-IR)
colors present tighter correlations with the absolute magnitude than
optical colors. Based on a matched sample of Galaxy
Evolution Explorer (GALEX) and SDSS, Wyder et al. (2007) and
Schiminovich et al (2007) explored the ultraviolet-optical
(UV-optical) CMRs for both disk and bulge-dominated galaxies. They also
found that disk-dominated galaxies show a tight correlation between
specific star formation rate ($SFR/M_*$) and galactic stellar mass
($M_*$), which is denoted as the mass-SFR relation for convenience.  

On the other hand, interpretation of these observed CMRs of disk
galaxies is still an open question. van den Bosch (2002) and 
Bell et al. (2003) found that, contrary to the observed trends, 
their semi-analytic models (SAMs) predicted that faint spirals 
should be slightly redder than bright spirals. By introducing 
additional mechanics like active galactic nucleus (AGN) feedback, 
the more recent SAMs can predict positive slopes of the CMRs 
of spirals, i.e., lower mass systems being bluer than massive systems 
(e.g. Kang et al. 2005, Croton et al. 2006). However, the slopes 
of the predicted CMRs are still shallower than that of the observations 
(Kang et al. 2005). This is mainly due to the fact
that it is difficult for the SAMs to build a scenario that 
permits low mass spiral to be averagely younger 
than high mass system in the hierarchical cosmogonies.

Parametrized modelling, which adopts phenomenological descriptions 
for some complicated processes, is another fruitful tool to 
explore the galactic formation and evolution (Tinsley 1980). 
In fact, parametric models have been successfully applied to
investigate the formation and evolution of the Milky Way disk (Chang
et al. 1999; Chiappini et al. 2001; Renda et al. 2005; Naab \&
Ostriker 2006; Fu et al. 2009; Matteucci et al., 2009) and other
nearby disk galaxies (Prantzos\& Boissier 2000; Molla \& Diaz 2005;
Dalcanton 2007; Yin et al. 2009). Besides these specific galaxies,
Boissier \& Prantzos (2000, hereafter BP00) extended their
calibrated model on the Milky Way disk to study the chemical and
spectro-photometric evolution of disk galaxies and compared their
model predictions with the observed CMRs of de Jong (1996). They
found satisfactory agreement with local observations provided that the
star formation time-scales of low-mass galaxies are longer than
their high-mass counterparts. However, their model predictions 
have not been compared with observations at higher redshift 
due to the limited observations at that time.

Recently, several new scaling relations of disk galaxies are 
statistically found and huge amount of observed data are available, 
not only in the local universe but
also for intermediate and high redshift galaxies.
Motivated by the work of BP00, we try to construct a new 
parametric model to explain the observed correlations in both local 
and intermediate redshift universe. Our results show that 
the mass-SFR relations at intermediate redshifts provide
tight constraints on the modelling of gas cooling history. 
Our results suggest that the Gaussian infall rate may be more 
reasonable to describe the gas cooling process than the 
exponential infall rate, especially for low-mass systems. 
 
The outline of this paper is as follows. In Section 2, we describe
main assumptions and ingredients of our model in details.
Comparisons between model predictions and observations and our main
results are given in Section 3. Discussions are presented in Section
4 and the last section summarizes our main conclusions.

\section{The model}

We assume that the disk galaxy has been embedded in a dark matter
halo and primordial gas within the dark halo cools down to form a
rotationally supported disk. The disk is basically assumed to be sheet-like
and composed by  a set of independent rings, each 500pc wide, that
is, there is no gas radial flow between different rings. The details and essentials
 of our model are outlined in below sections.

\subsection{Present-day disk galaxy}

In the local Universe, galactic stellar mass $M_*$ is the
primary parameter that determines main properties of the galaxy
(Kauffmann et al. 2003a, 2003b), including the colors of disk galaxy
(Chang et al. 2006a).  In this paper, we assume $M_*$ is the primary 
parameter and all free parameters in our model are dependent on $M_*$, 
where $M_*$ is referred to the stellar mass of a galaxy at present 
time (i.e., z=0) if there is no specific notations.

For a present-day galaxy with galactic stellar mass $M_*$, we assume
the stellar mass surface density in the current time
$\Sigma_*(r,t_g)$ follows an exponential profile:
\begin{equation}
\Sigma_*(r,t_g)=\Sigma_*(0,t_g) \rm {exp}(-r/r_d),
\end{equation}
where $t_g$ is the cosmic age at the present time, which 
is 13.5 Gyr for the cosmogony adopted here (see following 
subsection). $\Sigma_*(0,t_g)$ is the central stellar mass 
surface density in the present-day and $r_d$ is the disc 
radial scale-length. Since a pure exponential disk has:
\begin{equation}
M_*=\Sigma_*(0,t_g) 2\pi r_d^2,
\end{equation}
$\Sigma_*(r,t_g)$ can be obtained if $M_*$ and the $M_* \sim r_d$
relation are given.

Shen et al. (2003) collected a large sample of SDSS galaxies and
obtained the mean $M_* \sim R_{50}$ relations for both early-type
and late-type galaxies in the local universe, where $R_{50}$ is the
aperture that contains $50\%$ of Petrosian flux (Stoughton et al.
2002). We adopt their observed $M_* \sim R_{50}$ relation of
late-type galaxies in z-band (equation 18 and table 1 of 
Shen et al. 2003) to determine $r_d (M_*)$ by adopting $r_d=R_{50}/1.68$ 
for a pure exponential disk, since z-band is a good indicator of 
the stellar mass.

\subsection{Gas infall rate}

Gas cooling is one of the key processes in modeling galactic
formation and evolution. Semi-analytical models, which have made large
successes in explaining a lot of observations (Kauffmann et al.
1993; Somerville \& Primack 1999; Cole et al. 2000; Kang et al.
2005; Croton et al. 2006; Fontanot et al. 2009), have presented a
method to describe gas cooling process. However, there are still
some open questions, for example, how to describe the mass and angular
momentum distributions of baryons in dark matter halo and their 
variations with time, and whether the specific angular momentum
conserves or not during gas cooling processes etc. (Efstathiou 2000;
Bullock et al. 2002; Chen et al. 2003). None of these questions 
has an answer clear enough to construct a simple model, at least 
at the present stage of our knowledge.

In this paper, we turn to another approach, i.e., the phenomenological 
approach, which adopts a parametric description for some 
complicated processes. Comparisons between model predictions and 
observations may suggest possible ranges for model
free parameters in the model and then help to know more information
about main properties of the processes described by these parameters. 
We assume that there is 1 Gyr time delay for the disk galaxy 
formation, which corresponds to $z \sim 6$ under the adopted 
cosmology in this paper. After that, the disk origins and grows 
by continuous gas infall from the dark halo and the metallicity 
of the infalling gas is primordial. Differing from BP00, who 
adopted an exponential infall rate, we follow Prantzos \& 
Silk (1998) and Chang et al. (1999) and adopt a Gaussian 
formula of gas infall rate. Comparison between exponential-infall 
model and Gaussian-infall model will be presented in Section 4. 

The Gaussian infall rate $f_{in}(r,t)$ (in units of
$\rm{M_{\odot}} \rm {pc}^{-2} \rm{Gyr}^{-1}$) is assumed to be:
\begin{equation}
f_{in}(r,t)= \frac{A(r)}{\sqrt{2\pi}\sigma}e^{-(t-t_p)^2/2\sigma^2}
\end{equation}
where $t_p$ is the characteristic time corresponding to the maximum 
gas infall rate (hereafter, we call $t_p$ as the infall-peak time) 
and $\sigma$ is the full width at half maximum of the peak. We adopt 
$\sigma=3Gyr$ throughout this paper to promise a relatively wide distribution 
of the infall rate. In fact, our calculations show that the varying of $\sigma$ 
in the range $2 \sim 4$ Gyr do not have large influence on the results. 
The $A(r)$ are a set of separate quantities determined by the boundary condition
$\Sigma_*(r,t_g)$. In practice, we iteratively obtain $A(r)$ by normalizing 
the resulted stellar mass surface density distribution at 
$t=t_g$ to be very close to its observed value $\Sigma_*(r,t_g)$. 

We should emphasize that the infall-peak time $t_p$ is 
the most important free parameter in our model that determines 
main properties of gas infall history. Indeed, $t_p$ regulates 
the shape of gas accretion history. For example, $t_p
\rightarrow 0$ corresponds to a time-declining infall rate, while
$t_p \rightarrow \infty$ corresponds to a time-increasing 
gas infall rate. Since we only focus on the global quantities 
of galaxy in this paper, we assume $t_p$ does not vary with 
radius $r$ but may be dependent on the galactic stellar mass $M_*$.

In this paper, we assume a cosmology with $\Omega_M=0.3,
\Omega_{\Lambda}=0.7$ and $H_0=70 \rm {km s^{-1} Mpc^{-1}}$.
Correspondingly, the cosmic time at z =0 is $t_g=13.5$
Gyr and the cosmic time at z =3 (0.4) is t = 2Gyr (9Gyr). 

\subsection{Star formation law and stellar evolution }


Star formation (SF) process is so complicated that it is very
difficult to fully understand the physical nature of star formation
regulation (Silk 1997, Kennicutt 1998a,b, Elmegreen 2002).
Fortunately, observations of galaxies on global scales reveal a
tight correlation between the average star formation rate per unit
area and the mean surface density of cold gas (e.g. Kennicutt
1998a,b). This observed relation is later called as Kennicutt SF
law. Recently, Kennicutt et al. (2007) found that, on spatial
scales extending down to at least 500 pc, the star formation rate 
(SFR) surface density also correlates with the local gas 
surface density, following a similar power law as that of 
the global quantities.

We adopt the Kennicutt SF law in this paper:
\begin{equation}
\Psi(r,t)=0.25\Sigma_{g}(r,t)^{1.4},
\end{equation}
where the SFR surface density $\Psi(r,t)$ is measured in units of
$\rm{M_{\odot}} \rm {pc}^{-2} \rm{Gyr}^{-1}$ and the cold gas surface density
$\Sigma_g(r,t)$ is in units of $\rm{M_{\odot}pc}^{-2}$.

Regarding the chemical evolution of the galactic disk, both
instantaneous-recycling approximation (IRA) and instantaneous mixing
of the gas with stellar ejecta is assumed, i.e., the gas in a fixed
ring is characterized by a unique composition at each epoch of time.
We take the classical set of equations of galactic chemical
evolution from Tinsley (1980) and adopt the return fraction  $R=0.3$
and stellar yield  $y=1Z_{\odot}$.

The updated stellar population synthesis (SPS)
model of Bruzual \& Charlot (2003), i.e., CB07, is adopted in the
present paper,  with the stellar IMF being
taken from Chabrier (2003). The lower and upper mass limits are
adopted to be $0.1 M_{\odot}$ and $100 M_{\odot}$, respectively.

\subsection{Gas outflow rate}

Gas outflow process may also influence the evolution of galactic 
disk. During the continuous star formation and evolution process, 
supernova (SN) explosions
eject part of energy into interstellar medium and some amount of gas
will be heated and expelled from the galactic disk if its kinematic energy
exceeds the bonding energy. Therefore, gas outflow rate may be
proportional to the SN explosion rate and then proportional to 
SFR surface density $\Psi(r,t)$ if the initial mass function (IMF)
does not vary with time. Furthermore, gas outflow rate may also 
correlate with galactic mass since low mass galaxies have shallower 
potential wells and then it is much easier for them to expel gas out 
than high mass systems (Kauffmann et al. 1993).

In this paper, we assume the outflowing gas has the same metallicity
as that of the interstellar medium at that time and will not fall 
again to the disk. The gas outflow rate $f_{\rm out}(r,t)$ (in 
units of $\rm{M_{\odot}} \rm {pc}^{-2} \rm{Gyr}^{-1}$) is 
assumed to be proportional to the SFR surface density $\Psi(r,t)$:
\begin{equation}
f_{\rm out}(r,t)=b_{\rm out}\Psi(r,t),
\end{equation}
where $b_{\rm out}$ is another free parameter in our model 
and may also be a function of galactic stellar mass $M_*$. 
Integrating the left side of the above equation, 
we derive $\int_{0}^{t_g} \int_{0}^{\infty} 
f_{\rm out}(r,t)\,{\rm d}r{\rm d}t =M_{\rm out}$, where 
$M_{\rm out}$ is the total mass expelled from the galaxy. 
Integrating the right side of Eq. 5 gets
$\int_{0}^{t_g} \int_{0}^{\infty} 
b_{\rm out}\Psi(r,t)\,{\rm d}r{\rm d}t =b_{\rm out}(1-R) M_*$, 
where $R$ is the return fraction. Then, we derive $M_{\rm out}=
 b_{\rm out}(1-R) M_*$. Therefore, $b_{\rm out}$ is proportional 
 to the fraction of total mass expelled relative to the final 
 mass locked in stars. 
      
In summary, our primary input parameter is the galactic stellar mass
$M_*$. Free parameters are infall-peak time $t_p$ and outflow 
parameter $b_{out}$. The combination of gas
infall rate and outflow rate determines the variation rate of the
total (gas+star) mass surface density. Assuming the initial total
surface density to be zero, we can numerically obtain the total mass
surface density at any time after free parameters being given. Adding
the SF law, we can describe how many cold gas turns into
stellar mass and then calculate the chemical and color evolution of
the disk. For simplicity, we assume that there is no scatter of the
model parameters $t_p$ and $b_{\rm out}$ at any given $M_*$.
In other words, our simple model only investigates the average
trend of the correlations between different quantities and does not
include their scatters.

\section{Model results versus observations}

Using the above model, for one galaxy at given $M_*$, we
independently calculate the evolution of cold gas in each ring along
the galactic disk. At each time step, we derive a set of individual
quantities, such as the SFR surface density, gas mass surface density, multi-band
luminosity and metallicity, and then integrate them to obtain the
global quantities for the whole galaxy. After that, our model
predictions can be directly compared with observations.

In this section, we present our main results step by step. Firstly,
we briefly describe the adopted observations to be compared with
model predictions. Then, we investigate the influence of free
parameters on model predictions, including not only $t_p$ and 
$b_{\rm out}$, but also the constance in the SF law. Finally,  
we construct a reasonable mass-dependent model and explore if 
its predictions can be consistent with the observed 
data.


\subsection{Observations}

Here, we show and summarize the observed mass-color relations, the 
mass-metallicity relation, mass-SFR relation and the mass-gas 
fraction relation for local galaxies, 
which are mainly taken from papers based on SDSS database. 
We also include the observed mass-SFR relations at redshift 
$z=0.3, 0.5, 0.7, 0.9 $.  

{\it \bf Mass-color relations:} Based on the matched sample of
SDSS and GALEX, Salim et al. (2007) and
Schiminovich et al. (2007) obtained the UV-optical color magnitude
diagrams of star forming galaxies in the local universe and 
investigated their physical properties as a function of specific
star formation rate ($SFR/M_*$) and stellar mass ($M_*$). We take 
the observed $(NUV-r) \sim M_* $ relations of
disk-dominated galaxies from Table 2 of Schiminovich et al. (2007)
and plot them in the left-upper panel of Fig. 1, where the open
circles are the data without dust-correction and the
squares denote those after dust corrections. The error-bars denote 
the dispersions of the data distributions.

Chang et al. (2006a) collected a large sample of late-type galaxies
from SDSS (DR4) and 2MASS to explore their CMRs in both optical and
optical-IR bands. The photometric aperture mismatch between SDSS and
2MASS is taken into account by correcting the SDSS magnitudes to the
isophotal circular radius, where the 2MASS magnitudes are measured,
based on the radial profile of surface brightness of SDSS galaxies. We
take the observed multi-band (optical and optical-IR) color-$M_*$
relations from Chang et al.(2006a) and plot them in Fig. 1,
where the square denotes the mean value of color after
dust-corrections and the error-bar represents the distribution
dispersion in each bin. Since the SPS models seem to be not
accurately reproducing the shape of the spectrum between 4000 and
5000 $\AA$, which causes an offset in g-r and r-i as the r
filter passes through this (rest-frame) wavelength region (Wake et
al. 2006), we ignore the r filter data and replace them with g
filter data in this paper.

{\it \bf Mass-metallicity relation:} The mass-metallicity relation 
is another tight constraint on models of galactic formation and
evolution. Based on the SDSS imaging and spectroscopy data, Tremonti
et al. (2004) derived gas-phase oxygen abundances and stellar masses
of a large sample of star-forming galaxies and found a tight
correlation between stellar mass and metallicty, spanning over 3
orders in stellar mass and a factor of 10 in metallicity.  Liang
et al. (2007) adopted another method to measure the gas-phase
metalllcity of a similar sample of star-forming galaxies from SDSS
(DR4). Within 27 stellar mass bins from $log(M_*/M_{\odot})=8.0$ to
$10.6$, they stacked the spectra of several hundreds (even
thousands) galaxies in each bin and derived their electron
temperature $T_e$ and hence obtained the $T_e$-based oxygen
abundance. In this paper, we adopt these two sets of observed
relations between gas phase metallicity $log(O/H)+12$ and stellar
mass $M_*$. The results are plotted in the right-second panel of
Fig. 1, where the data taken from Tremonti et al. (2004) are shown
by yellow shaded area and that taken from Liang et al. (2007) are
denoted by green filled circles with error bars.

{\it \bf Mass-SFR relation:} The specific star
formation rate $SFR/M_*$ is an important quantity describing the
star formation history. If the initial mass function does not vary
with time, stellar mass $M_*$ should be roughly proportional to mean
star formation rate during the whole history $<SFR>$, thus, $SFR/M_*
\propto SFR/<SFR>$. Therefore, a galaxy having higher $SFR/M_*$ means
larger fraction of its stars formed in later time. The observed
$SFR/M_* \sim M_*$ relations of star-forming galaxies in the
local universe are taken from Schiminovich et al. (2007) and plotted
in right-third panel of Fig. 1, where the SFRs taken from Brinchmann et
al. (2004) and from Salim et al. (2007) are denoted as the open
circles and squares, respectively.

Besides the local data, the observed mass-SFR relations at different 
redshift are also available. Zheng et al. (2007) measured the SFR 
from UV and IR luminosities for a sample of $\sim$15,000 galaxies 
selected from two COMBO-17 fields. Combining with the stellar mass 
of these galaxies derived by Borch et al. (2006), Zheng et al. (2007) 
derived the observed mass-SFR relations at four redshift bins 
$z=0.3, 0.5, 0.7, 0.9 $. We adopt their data and plot them as 
filled squares in Fig. 2.  

{\it \bf Mass-gas fraction relation:} The observed gas fraction is also 
important constraint on the model. Zhang et al. (2009) collected a sample of 
721 galaxies from cross-matching the SDSS (DR4) galaxies, the 2MASS XSC and 
the HyperLeda HI catalogue (Paturel et al. 2003) and obtained their observed gas 
fraction. The observed data of gas fraction are taken from
sample 1 of Zhang et al. (2009) and plotted as small dots in the right-fourth
panel of Fig. 1. In the same panel, the filled triangles represent the observed 
data of gas fraction and stellar mass for a sample of 60 local galaxies, which 
are taken from Table 1 of McGaugh (2005). 



\subsection{Influence of free parameters}

\begin{figure}[!t]
  \centering
\includegraphics[height=8cm,width=8cm]{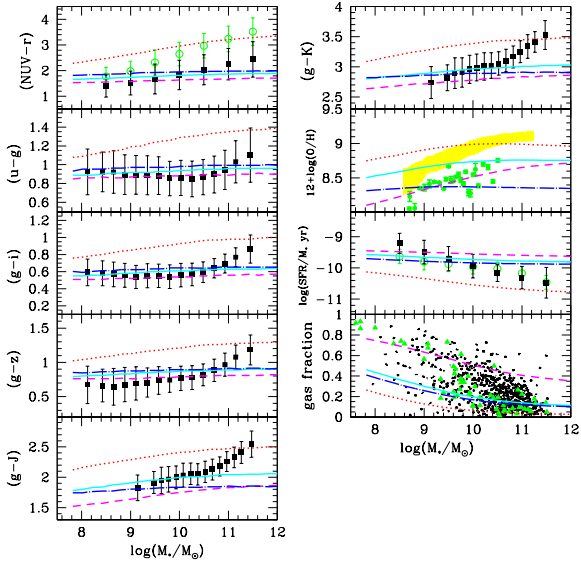} 
\caption{Comparisons between model predictions and local
observations. Different lines correspond to various
parameter groups: dot lines ($t_p, b_{out}$) =
(5Gyr,0), solid lines ($t_p, b_{out}$) = (10Gyr,0),
dot-dash lines ($t_p, b_{out}$) = (10Gyr,1). Dash
lines adopts ($t_p, b_{out}$) = (10Gyr,0), but a modified 
star formation law $\Psi(r,t)=0.05\Sigma_{g}^{1.4}(r,t)$ is 
assumed to reduce the star formation efficiency. The observed 
data are described in Section 3.1 in details. }
\end{figure}

\begin{figure} [!t]
  \centering
  \includegraphics[height=8cm,width=8cm]{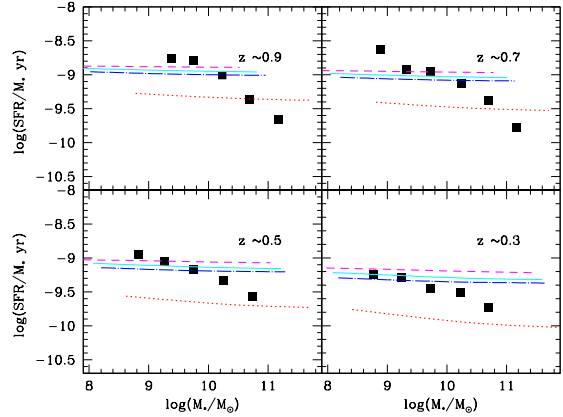}
\caption{Model predictions compare to the observed data at 
redshift $z \sim 0.3,0.5, 0.7, 0.9,$ respectively. The notations of 
different lines are the same as Fig.1. The observed data are taken
from Zheng et al.(2007).  }
\end{figure}

In this section, we show model predicted scaling relations with
different settings of free parameters and compare them with the
observations. Fig. 1 compares model predictions with local 
observations. Fig. 2 plots the observed and model predicted 
mass-SFR relation at $z \sim 0.3,0.5, 0.7, 0.9,$ respectively.
The observed data in Fig. 1 and 2 are described in details in
subsection 3.1. We emphasize that, for both observations
and model predictions, $M_*(t)$ in Fig.2 refers to the galactic
stellar mass at given redshift (or evolutionary time step).

We start from a prior model adopting $t_p=10$Gyr and $b_{out}=0$, 
which corresponds to a late infall-peak time and no gas outflow is 
considered. The results of our prior model are plotted as solid lines 
in Fig.1 and 2. It should be pointed out that the
construction of this prior model is arbitrary and far from a
reasonable description of the formation of any galactic disk. This 
setting only aims to provide a benchmark to be compared with other model 
predictions and then investigate the influences of different 
settings of free parameters on model results.

Firstly, we fix $b_{out}=0$ and change $t_p$ from 10Gyr
to 5Gyr, i.e. replacing the prior model with a shorter infall-peak
time. The model predictions are shown as the dot lines in
Fig. 1 and 2.  It can be seen that the model adopting later $t_p$ 
predicts bluer color, lower metallicity, higher gas fraction and 
higher specific star formation rate $SFR/M_*$ both in the local 
Universe and at intermediate redshift, at least up to $z \sim 1$. 
This is mainly due to the fact that, in our model, the setting of 
later infall-peak time corresponds to slower gas infall 
process and thus slower star formation processes. Furthermore, 
Fig. 1 and 2 show that the separations between the solid and 
dash lines almost cover the observed range of different 
data sets, especially the $SFR/M_*$ at intermediate redshifts. 
In other words, the infall-peak time $t_p$ significantly influence 
the galactic evolution.  

The impact of outflow (described by $b_{\rm out}$) on the model
predictions is presented by the dot-dash lines in Fig. 1 and 2, 
where we adopt $b_{\rm out}=1.0$ and $t_p=10$Gyr. Comparison 
between solid lines and dot-dash lines shows that the gas outflow 
process mainly influences the final gas-phase metallicity, since it 
takes a fraction of metal away from the disk.
Moreover, after including gas outflow process, the model predicts
bluer optical-IR (both $g-J$ and $g-K$) color, while the impact on
the optical color is very small, if there is any. This is mainly
due to the fact that the stellar mean age influences the optical
color more strongly than the optical-IR color, while the mean
stellar metallicity influences the optical-IR color more strongly
than the optical color (Chang et al. 2006b).

\begin{figure} [!t]
  \centering
  \includegraphics[height=8cm,width=8cm]{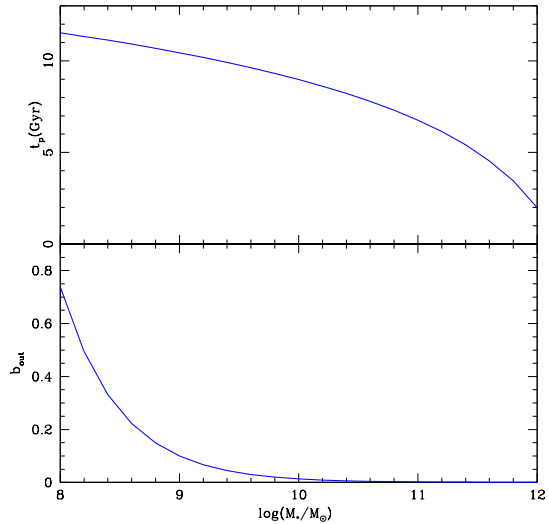}\\
\caption{The adopted infall-peak time $t_p(M_*)$ and 
outflow parameter $b_{out}(M_*)$ in our mass-dependent model. }
\end{figure}

The star formation law is another important ingredient in our model. 
In this paper, we adopt a Kennicutt star formation law. Although 
it works well in the region of high surface density, we should 
be caution to apply it to low density region, such as the 
outskirts of disk galaxies. To explore the influence of the adopted 
star formation law on our model predictions, we adopt a modified 
star formation law $\Psi(r,t)=0.05\Sigma_{g}^{1.4}(r,t)$ to reduce 
the star formation efficiency (SFE) and plot the model predictions as 
dash lines in Fig. 1 and 2. The free parameter combination is adopted 
to be $(t_p, b_{out}$) = (10Gyr,0). It shows that the low SFE results 
in bluer color, lower metallicity and higher $SFR/M_*$. The most 
outstanding point appears in the mass-gas fraction relation. The 
model adotping the low SFE predicts much higher gas fraction than that of 
standard SFE. In other words, the predicted gas fraction is very sensitive 
to the adopted SFE. However, except the panel of gas fraction, the separations 
between the solid and dash lines are very small compared to the observed range 
of different data sets, i.e., the variation of SFE may not dramatically change our model 
predictions. Considering the fact that there still 
lacks an alternative choice more reasonable than the Kennicutt star 
formation law, we choose to adopt the Kennicutt star formation 
law throughout this paper if there is no specific statement.  

\subsection{A mass-dependent model}

\begin{figure}[!t]
  \centering
  \includegraphics[height=8cm,width=8cm]{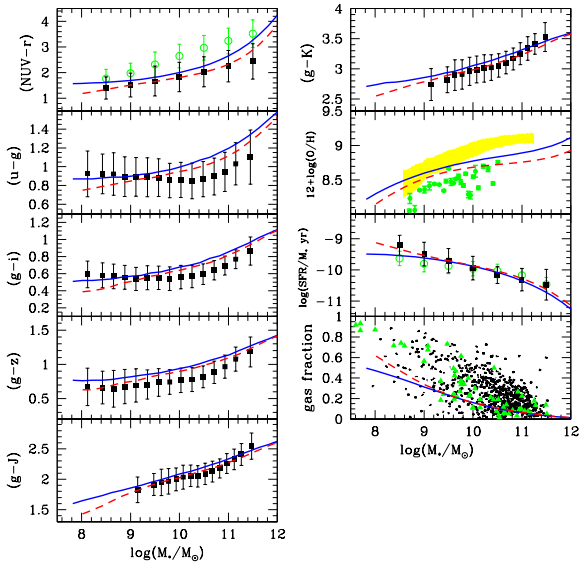}\\
\caption{Our mass-dependent model predictions (solid lines) compare to 
the local observations. The dash lines show the predictions of
the exponential-infall model, which 
is described in Section 4 in details.}
\end{figure}

\begin{figure} [!t]
  \centering
  \includegraphics[height=8cm,width=8cm]{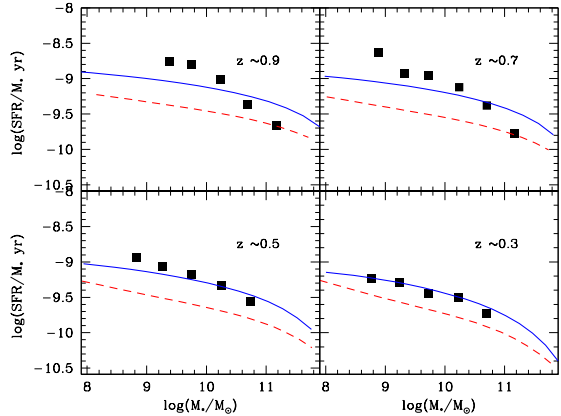}\\
\caption{Comparison between mass-dependent model predictions and
observations at intermediate redshifts. The notation of solid and 
dash lines are the same as Fig. 4.  }
\end{figure}

Model calculations in above section show that galaxies form by
later infall-peak time will result in bluer color and higher 
specific star formation rate, while gas outflow
process will decrease the gas metallicity in the present day. The local
observations show that less massive disk galaxies have bluer color,
lower metallicity when compared with their high mass counterparts
(see Fig. 1). Therefore, it is naturally to assume that low mass
disk galaxies may have later infall-peak time and larger
outflow parameter than massive systems.

Motivated by the above discussions, we try to construct a model
in which the infall-peak time $t_p$ and outflow parameter $b_{out}$ 
are only dependent on galactic stellar mass $M_*$. The adopted 
$t_p(M_*)$ and $b_{out}(M_*)$ are plotted in Fig. 3. The adoption 
considers the facts below.
Since massive spirals have very old stellar population and their
potential wells are so deep that there is almost no outflow occurs,
we set one boundary condition at high-mass end 
$logM_*/M_{\odot}=12$ to be $t_p=2\rm{Gyr}$ and $b_{\rm out}=0$. 
Regarding the facts that low mass spirals
are observed to have blue colors and their potential well is shallow
and then should have large outflow rate, another boundary condition
of $t_p=11.5\rm{Gyr}$ and $b_{\rm out}=0.7$ 
is chosen at the low-mass end $logM_*/M_{\odot}=8$.

Moreover, the general trends of $t_p(M_*)$ and $b_{out}(M_*)$ 
are consistent with
previous studies of disk galaxies in the local group. For example,
the Milky Way disk have stellar mass about
$logM_*/M_{\odot} = 10.6$ (Yin et al. 2009). Our mass-dependent model 
adopts the corresponding infall-peak time about
7.8 Gyr, which is roughly in agreement with our previous chemical
evolution models (see Chang et al. 1999 and Yin et al. 2009 for
details). For M33, chemical evolution models show that this low mass
disk galaxy should form by slow accretion process (Wang et al. 2009;
Magrini et al. 2007), which equals to a late infall-peak time in our model .


The mass-dependent model predictions are plotted as solid 
lines in Fig. 4 and 5. As expected, under the
assumption that $t_p$ and $b_{out}$ all decrease with
the increase of $M_*$, our model do predict mean trends of most of
observations both in local Universe and at different redshift, at 
least to $z\sim 1$. It should be emphasized that, although 
the accurate values of free parameters in our model are not 
unique for any individual galaxy, the main trend that both $t_p$ 
and $b_{out}$ decrease with the increase of $M_*$ are robust. 

\section{Discussions: exponential-infall model }

In above section, we adopt the Gaussian infall rate. However, 
the exponential infall rate is much more widely assumed in previous models.
In this section, firstly, we will compare our mass-dependent model 
assuming the Gaussian infall rate with that adopting the exponential 
infall rate. We hope the observed data may provide constrains on the form 
of gas infall rate and then help us to know more about 
the gas cooling process in disk galaxies. Secondly, we will discuss 
the difference of our model with that of BP00.  

In order to investigate the possible influence of the infall form 
on the galactic evolution, we fix other ingredients of the mass-dependent 
model but replace the Gaussian infall rate by an exponential one:
\begin{equation}
g_{in}(r,t)= B(r)e^{-t/\tau},
\end{equation}
where $g_{in}(r,t)$ is the gas infall rate in units of
$\rm{M_{\odot}} \rm {pc}^{-2} \rm{Gyr}^{-1}$, $\tau$ 
is the infall time-scale. The $B(r)$, which is similar as 
$A(r)$ in Eq. 3, is a set of 
normalization quantities constrained by the 
boundary condition $\Sigma_*(r,t_g)$.

\begin{figure} [!t]
  \centering
\includegraphics[height=8cm,width=8cm]
{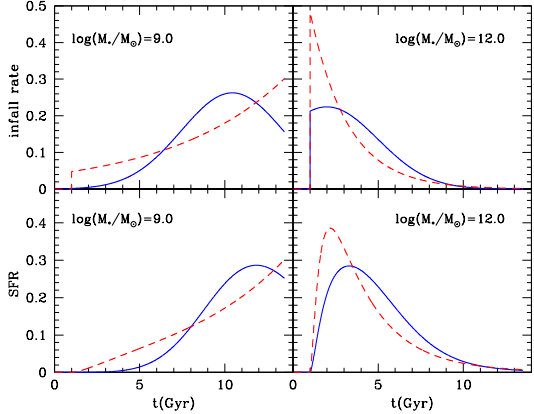}
\caption{Gas infall history and total star formation history of 
galaxies of $\rm{logM_*}/\rm{M}_{\odot}$ = 9.0 and  
$\rm{logM_*}/\rm{M}_{\odot}$ = 12.0. The y-axis of two 
left panels are normalized by $10^{9}\rm{M}_{\odot}\rm{Gyr}^{-1}$, 
while that of two right panles are normalized by $10^{12}\rm{M}_{\odot}
\rm{Gyr}^{-1}$.  The solid and dash lines 
represent the model results adopting Gaussian and exponential 
infall rate, respectively.  }
\end{figure}

Accordingly, we also assume that infall time-scale $\tau$ 
is only dependent on the stellar mass. To be consistent 
with the observed data, we adopt the form of $\tau(M_*)$ as:
\begin{equation}\label{eq:tau}
\frac{\tau}{\rm{Gyr}} = \frac{16}{e^{(\rm {log}M_*/M_{\odot}-10)}
-e^{-(\rm {log}M_*/M_{\odot}-10)}},
\end{equation} 
which gives that the infall time-scale of massive galaxy 
($M_*>10^{10}\rm{M}_{\odot}$) is positive and decreases with the increase of $M_*$, 
while the infall time-scale of low-mass galaxy ($M_*<10^{10}\rm{M}_{\odot}$) 
is negative and increases with the decrease of $M_*$. It should 
be pointed out that, although $\tau(M_*)$ is discontinue at 
$M_*=10^{10}\rm{M}_{\odot}$, the gas infall history described by 
the adopted $\tau(M_*)$ varies smoothly from increasing with 
time to being constant and then to decreasing with time as galaxy 
changes from low-mass to intermediate and then to high-mass systems.   

To visually compare the Gaussian and exponential infall rate, 
Fig. 6 plots the gas infall history and total star formation history 
of two galaxies $M_*=10^{9}\rm{M}_{\odot}$ and 
$M_*=10^{12}\rm{M}_{\odot}$, which are normalized 
by $10^{9}\rm{M}_{\odot}\rm{Gyr}^{-1}$ (left two panels) and 
 $10^{12}\rm{M}_{\odot}\rm{Gyr}^{-1}$ (right two panels), 
 respectively. The solid and dash lines 
represent the model results adopting Gaussian and exponential 
infall rate. According to the assumption that there is 1Gyr delay 
for the halo formation, all of the gas infall throughout this paper 
begin at $t=1\rm{Gyr}$. It shows that, comparing with the star 
formation rate predicted by Gaussian-infall model, the exponential-infall 
model predicts higher SFR at early stage and lower SFR later. This point 
directly results in the differences of the mass-SFR relations at 
intermediate redshifts predicted by these two models (see following discussions).

In order to compare with the observations, we plot the predictions of 
the exponential-infall model as dash lines in Fig. 4 and 5. It can be seen that, 
although both the solid and dash lines can be roughly consistent 
with the local observations, the exponential-infall model does predict 
lower $SFR/M_*$ than observations at intermediate redshifts. Since 
the stellar population of local galaxy is resulted along its whole 
star formation history, local observations do not have the ability to 
distinguish whether the model is viable at different time steps. Fortunately, 
the observed data at intermediate redshifts are now available and can 
provide tighter constraints on the model of galactic evolution. In other 
words, our results suggest that the Gaussian infall rate may be more 
reasonable to describe the gas cooling process than the exponential infall rate. 

In fact, by fitting synthetic energy distribution to the multi-band 
photometry of a sample of actively star-forming galaxies at $z\sim2$, 
Maraston et al. (2010) concluded that adopting an exponentially increasing 
SFRs and a high starting redshift appear to provide more reasonable 
results than that of an exponentially declining SFR in the case 
of star forming galaxies at redshift beyond $\sim1$. Unfortunately, 
both their model and ours are phenomenological models. It is still 
beyond the scope of this paper to answer why the efficiency of gas cooling process 
or/and star formation process may be depressed at early stage and 
why the infall-peak time of low-mass galaxy is later than 
high-mass systems.

Indeed, the semi-analytical models and studies using hydrodynamical 
simulations may provide more physical and reasonable explanation on 
how gas accreted onto the galaxies than phenomenological models. However, 
there is still a long way to go to fully understand the gas cooling processes. 
For example, the semi-analytical models assumes that gas within galaxy 
halos initially reaches the temperature of the virial halo, while  
some studies using high-resolution cosmological hydrodynamical simulations found that 
the majority of baryons may never reach the virial temperature as they 
accreted onto galaxies (Keres et al. 2005; Dekel $\&$ Birboim 2006). 
Brooks et al. (2009) found that low-mass galaxies 
are instead dominated by accretion of gas that stays well below the viral 
temperature and even for galaxies at higher masses, cold flows dominate the 
growth of the disk at early times. All these findings called the general 
virial temperature assumption into question, but may give more information 
to improve the semi-analytical models and help us to understand 
more clearly on how the gas gets into the galaxies. 
  
Finally, comparing our model to that of BP00, the main difference is that 
BP00 adopted the exponential infall rate while we adopt 
the Gaussian infall rate. BP00 also notice the fact that, if the 
exponential declining infall rate is assumed, the predicted 
stellar population will be too old to reproduce 
the observed blue colors of low-mass galaxies even if the 
infall time-scale tends to be as large as $\tau\rightarrow\infty$. 
They assumed a negative infall time-scale for the most extended and
less massive galaxies in their model, i.e. the infall rate 
increases with time. However, according to our discussions above, 
we guess that the model of BP00 may predict lower $SFR/M_*$ at 
intermediate redshift than the observed data although their 
model predictions can be in good agreement with the 
local observations.    



\section{Summary}

In this paper, we present a phenomenological model for the formation and
evolution of disk galaxies. It is assumed that the disk origins and
grows by gas infall. We adopt the Gaussian form of gas infall rate 
with one free parameter, infall-peak time $t_p$. We also include 
the contribution of gas outflow induced by supernova feedback. 
The Kennicutt star formation law is adopted to describe how much 
cold gas turns into stellar mass. We calculate the evolution of disk
galaxies and compare model predictions with observations.  We
find that:

\begin{enumerate}
\item Model predictions are very sensitive to the 
infall-peak time $t_p$ in that the model adopting later $t_p$ 
results in blue colors, low metallicity, high specific star 
formation rate and high gas fraction, while gas outflow rate mainly
influences the gas-phase metallicity and star formation efficiency mainly 
influences the gas fraction.

\item We construct a mass-dependent model by
assuming that low mass galaxy has later $t_p$ and larger $b_{out}$
than massive systems. It is shown that our mode predictions
are in fairly agree with not only the local observations, but also 
the observed mass-SFR relations at intermediate redshifts.

\item It shows that, comparing with the star formation rate 
predicted by the Gaussian-infall model, the exponential-infall 
model predicts higher star formation rate at early stage 
and lower star formation rate later. Our results suggest that 
the Gaussian infall rate may be more reasonable to describe the gas 
cooling process than the exponential infall rate, especially for low-mass systems. 
\end{enumerate}

We emphasize that, here we present a phenomenological
model. Through comparison between model predictions and
observations, our main aim is to investigate main properties of the
formation and evolution of disk galaxy, especially the trend as a 
function of galactic stellar mass. Unfortunately, there is still 
a long way to go to fully understand why low-mass galaxy may 
have later infall-peak time than high-mass systems 
and why the Gaussian infall rate may be more prefered than 
the exponential infall rate.

\begin{acknowledgements}

We thank the anonymous referee for helpful comments. This work is 
supported by the National Science Foundation of China
No.10573028, 10403008, the Key Project No. 10833005 and No. 10878003, 
the Group Innovation Project No.10821302, by 973 program No. 
2007CB815402.

\end{acknowledgements}

\end{document}